\documentstyle[12pt]{article}
\def\bqn{\begin{equation}}
\def\nqn{\end{equation}}

\newtheorem{defi}{Definition}[section]
\newtheorem{lem}[defi]{Lemma}
\newtheorem{prop}[defi]{Proposition}
\newtheorem{theo}[defi]{Theorem}
\newtheorem{rem}[defi]{Remark}

\newcommand{\p}{\partial}

\def\C{{\rm\kern.24em \vrule width.02em height1.4ex depth-.05ex \kern-.26em
C}}
\def\R{{\rm I\kern-.20em R}}
\def\Z{{\rm\kern.26em \vrule width.02em height0.5ex depth0ex \kern.04em
\vrule  width.02em height1.47ex depth-1ex \kern-.34em Z}}
\def\Q{{\rm\kern.24em \vrule width.02em height1.4ex depth-.05ex \kern-.26em
Q}}
 
\begin{document}

\vspace{2.5cm}
\begin{center}{\large \vspace{4.0cm} 
\bf Volume preserving multidimensional integrable 
 systems and Nambu-Poisson Geometry}\\
 \vspace{1cm}
Partha Guha \\
S. N. Bose National Centre for Basic Sciences \\ JD Block, Sector-3, Salt
Lake City \\
Calcutta - 700091, INDIA  \\ and \\ Mathematics Institute,
 University of Warwick\\ Coventry CV4 7AL, ENGLAND. \\

\vspace{3cm}

\noindent{\sl dedicated to the memory of Dr. B.C. Guha}

\vspace{3cm}

\begin{abstract}
In this paper we study generalized classes of volume preserving 
multidimensional integrable systems via Nambu-Poisson mechanics. 
These integrable systems belong to the same class of dispersionless 
KP type equation. Hence they bear a close resemblance 
to the self dual Einstein equation.
 Recently Takasaki-Takebe provided the twistor  
construction of dispersionless
KP and dToda type equations by using the Gindikin's pencil of two forms.
In this paper we generalize this twistor construction to our systems.

\vspace{2cm}

\noindent{\bf Mathematics Subject Classification (1991)} 70H99, 58F07

\end{abstract}

\newpage

\end{center}

\section{Introduction}

In this article we study volume preserving diffeomorphic integrable 
hierarchy of three flows [1]. This is different from the usual two flows 
cases and this can be studied via Nambu-Poisson geometry. It is already known 
that a group of volume preserving diffeomorphisms in three dimension plays a 
key role in an Einstein-Maxwell theory whose Weyl curvature is self-dual and 
whose Maxwell tensor has an algebraically anti self-dual. Later Takasaki [2] 
explicitly showed how volume preserving diffeomorphisms arises in integrable 
deformations of self-dual gravity.

Nambu mechanics is a generalization of classical Hamiltonian mechanics, 
introduced by Yoichiro Nambu [3]. At the begining he wanted to formulate a 
statistical mechanics on ${\bf R}^3$, emphasizing that the 
only feature of Hamiltonian mechanics one should preserve 
is the Liouville theorem. He considered the following equations of motion
$$ \frac{d{\bf r}}{dt} ~=~ \nabla u({\bf r})  \wedge 
\nabla v({\bf r}), \hbox{   } \mbox{   } {\bf r} ~=~ (x,y,z) \in {\bf R}, $$
where $ x,y,z$ are dynamical variables and $u,v$ are two functions of
 ${\bf r}$. Then Liouville theorem follows from the identity
$$ \nabla \cdot (\nabla u({\bf r}) \wedge \nabla v({\bf r})) ~=~ 0.$$

He further observed that the above equation of motion can be cast into
$$ \frac{df}{dt} ~=~ \frac{\p (f,u,v)}{\p (x,y,z)},$$ 
the Jacobian of right hand side can be interpreted as a generalized Poisson
 bracket.

Hence the binary operation of Poisson bracket of Hamiltonian mechanics is
 generalized to $n$-ary operation in Nambu mechanics. Recently Takhtajan [4,5]
 has formulated its basic principles in an invariant geometrical 
 form similar to that of Hamiltonian mechanics. 

In this paper we shall use Nambu mechanics to study generalized 
volume preserving diffeomorphic integrable hierarchy. 
These classes of integrable systems are closely related to the self dual
Einstein equation, dispersionless KP
equations etc. In fact we obtain a higher dimensional analogue of 
all these systems. It turns out that all these systems can be 
written in the following form:
$$ d\Omega^{(n)} ~=~ 0$$
$$ \Omega^{(n)} \wedge \Omega^{(n)} ~=~ 0.$$
For $n~=~2$ we obtain all the self dual Einstein and dKP type equations.

Hence we obtain a common structure behind all these integrable system, so
there is a consistent and coherent way to describe all these systems. 
Here we unify all these classes of integrable systems by Gindikin's pencil 
or bundle of forms and Riemann-Hilbert problem ( twistor description).
Gindikin introduced these technique to study the geometry of the solution of
self dual Einstein equations. Later Takasaki-Takebe [6,7] applied
it to dispersionless KP and Toda equations.

\bigskip

This paper is organized as follows:\\
In section 2 we give some working definitions of Nambu-Poisson geometry.
In section 3 we construct our volume preserving integrable systems via
Nambu-Poisson geometry. If one carefully analyse these set of equations
then one must admit that they bear a close resemblance to the {\sl volume 
preserving KP equation}, so far nobody knows about this equation. It is
known that area preserving KP hierarchy (= dispersionless KP hierarchy)
plays an important role in topological minimal models ( Landau-Ginzburg
description of the A-type minimal models ). So we expect volume preserving
KP hierarchy may play a big role in low dimensional quantum field theories.  
Section 4 is dedicated to the twistor construction
of these systems.

I would like to end this introduction by expressing some regrets.
I apologize to the mathematicians that the presentation of this
paper is not regorous, simply because, I want to show the application
of the Nambu-Poisson geometry to the large spectrum of readers.

\section{Nambu-Poisson Manifolds}

In this section we will state some basic results of Nambu-Poisson 
manifold from the paper of Takhtajan [4].

Let ${\cal M}$ denote a smooth finite dimensional manifold and 
$C^{\infty }({\cal M})$ the algebra of infinitely differentiable 
real valued functions on ${\cal M}$. Recall that [4,5] 
${\cal M}$ is called a Nambu-Poisson manifold if there exits a 
$\R$-multi-linear map
\bqn
\{ ~,\ldots,~ \}~:~[C^{\infty }({\cal M})]^{\otimes n} \rightarrow 
C^{\infty }({\cal M})
\nqn
called a Nambu bracket of order $n$ such that
 $\forall f_1 , f_2 , \ldots , f_{2n-1} \in C^{\infty }({\cal M})$,
\bqn
\{ f_1, \ldots ,f_n \}=(-1)^{\epsilon(\sigma)}\{ f_{\sigma(1)}, \ldots ,
f_{\sigma(n)} \},
\nqn
\bqn
\{ f_1 f_2, f_3, \ldots ,f_{n+1} \}=
f_1 \{f_2, f_3, \ldots , f_{n+1} \} +
\{ f_1, f_3, \ldots, f_{n+1} \} f_2,
\nqn
and
\begin{eqnarray}
\{ \{ f_1, \ldots , f_{n-1}, f_n \}, f_{n+1}, \ldots, f_{2n-1} \} +
\{ f_n, \{ f_1, \ldots, f_{n-1}, f_{n+1} \}, f_{n+2}, \ldots , f_{2n-1} \} \\
 +  \cdots + \{ f_n, \ldots ,f_{2n-2}, \{ f_1, \ldots , f_{n-1}, f_{2n-1} \}\}
 =  \{ f_1, \ldots , f_{n-1}, \{ f_n, \ldots , f_{2n-1} \}\}, \nonumber
\end{eqnarray}
where $\sigma \in S_n$---the symmetric group of $n$ elements---and 
$\epsilon(\sigma)$ is its parity.  Equations (2) and (3) are the standard 
skew-symmetry and derivation properties found for the ordinary ($n=2$) Poisson 
bracket, whereas (4) is a generalization of the Jacobi identity and was called 
in [2] the fundamental identity. When $n~=~3$ this fundamental identity reduces to

\bigskip

\begin{eqnarray}
\{ f_1, f_2, f_3 \}, f_4, f_5 \} + \{ f_3, \{ f_1, f_2, f_4 \}, f_5 \} + \{ f_3, f_4, \{ f_1, f_2, f_5 \} \} \\
  =  \{ f_1, f_2, \{ f_3, f_4, f_5 \} \}. \nonumber
\end{eqnarray} 

It is also shown in [2] that Nambu dynamics on a Nambu-Poisson phase space involves $n-1$ so-called Nambu-Hamiltonians $H_1, \ldots, H_{n-1} \in  
C^{\infty }({\cal M})$ and is governed by the following equations of motion
\bqn
\frac {df}{dt} =  \{ f , H_1 ,\ldots,H_{n-1} \},~\forall f \in
C^{\infty }({\cal M}).
\nqn

A solution to the Nambu-Hamilton equations of motion produces an evolution 
operator $U_t$ which by virtue of the fundamental identity preserves  
the Nambu bracket structure on $C^{\infty }(M)$.

\begin{defi}
$ f \in C^{\infty}({\cal M})$ is called an integral of motion for the system if it satisfies
$$ \{ f,H_1,H_2, \cdots ,H_{n-1} \} ~=~ 0.$$
\end{defi}

Like the Poisson bivector the Nambu bracket is geometrically realized by a Nambu polyvector $\eta \in 
\Gamma (\wedge ^n TM)$, a section of $\wedge ^n TM$, such that 
\bqn
\{ f_1,\ldots,f_n \} = \eta (df_1,\ldots,df_n),
\nqn
which in local coordinates $(x_1,\ldots,x_n)$ is given by
\bqn
 \eta = \eta_{i_1 \ldots i_n}(x) \frac{\partial}{\partial x_{i_1}} \wedge \cdots
\wedge \frac{\partial}{\partial x_{i_n}},
\nqn
where summation over repeated indices are assumed.

It was stated in [4] that the fundamental identity (4) is equivalent to 
the following algebraic and differential constraints on the Nambu tensor
$\eta_{i_1 \ldots i_n}(x)$:
\bqn
S_{ij} + P(S)_{ij} = 0,
\nqn
for all multi-indices $i=\{i_1,\ldots,i_n \}$ and
$j=\{j_1,\ldots,j_n \}$ from the set $\{1,\ldots,N \}$,
where
\bqn
S_{ij} = \eta _{i_1 \ldots i_n} \eta _{j_1 \ldots j_n}
+ \eta _{j_n i_1 i_3 \ldots i_n } \eta _{j_1 \ldots j_{n-1} i_2}
+ \cdots + \eta _{j_n i_2 \ldots i_{n-1} i_1 } \eta _{j_1 \ldots j_{n-1} i_n}  
- \eta _{j_n i_2 \ldots i_n } \eta _{j_1 \ldots j_{n-1} i_1},
\nqn
and $P$ is the permutation operator which interchanges the indices $i_1$ and 
$j_1$ of $2n$-tensor $S$, and
\begin{eqnarray}
{} & {} & \sum_{l=1}^{N}  \left ( \eta_{l i_2 \ldots i_n}
\frac{\partial \eta_{j_1 \ldots j_n}}
{\partial x_l} + \eta_{j_n l i_3 \ldots i_n}\frac{\partial \eta_{j_1 \ldots
j_{n-1} i_2}}{\partial x_l} + \ldots + \eta_{j_n i_2 \ldots i_{n-1}l}
\frac{\partial\eta_{j_1 \ldots j_{n-1} i_n}}{\partial x_l} 
\right) \nonumber \\
& = & \sum_{l=1}^{N}  \eta_{j_1 j_2 \ldots j_{n-1} l} 
\frac{\partial \eta_{j_n i_2  \ldots i_n}}
{\partial x_l},
\end{eqnarray}
for all $i_2,\ldots,i_n,~j_1,\ldots,j_n = 1,\ldots,N$.

It was conjectured in [2] that the equation $S_{ij}=0$ is equivalent 
to the condition that $n$-tensor $\eta$ is decomposable, 
recently this has been proved by Alekseevsky and author[8], 
and more recently by Marmo et. al. [9] and many others,
so that any decomposable element in $\wedge^nV$, where $V$ 
is an $N$-dimensional vector space over 
$\R$, endows $V$ with the structure of a Nambu-Poisson manifold.

\bigskip

{\bf Example}

Let us illustrate how Nambu-Poisson mechanics works in practise. 
The example is the motion of a rigid body with a torque about the 
major axis introduced by Bloch and Marsden [10].

Euler's equation for the rigid body with a single torque $u$ about its major 
axis is given by
\bqn {\dot m}_1 ~=~ a_1 m_2 m_3 \nqn
\bqn {\dot m}_2 ~=~ a_2 m_1 m_3 \nqn 
\bqn {\dot m}_3 ~=~ a_3 m_1 m_2 + u \nqn
where $ u ~=~ -km_1 m_2$ is the feedback, 
$ a_1 ~=~ \frac{1}{I_2} - \frac{1}{I_3}$,
$ a_2 ~=~ \frac{1}{I_3} - \frac{1}{I_1}$ and 
$ a_3 ~=~ \frac{1}{I_1} - \frac{1}{I_2}$. We assume $ I_1 < I_2 < I_3$.

These equations can be easily recast into generalized 
Nambu-Hamiltonian equations of motion: 
\bqn \frac{dm_i}{dt} ~=~ \{ H_1, H_2, m_i \}, \nqn
where these equations involve two Hamiltonians and these are
    $$  H_1 ~=~ \frac{1}{2} ( a_2 m_{1}^2 - a_1 m_{2}^2 ) $$
    $$  H_2 ~=~ \frac{1}{2} ( \frac{a_3 - k}{a_1}m_{1}^2 - m_{3}^2 ). $$

\bigskip

When $$ a_1 ~=~ a_2 ~=~ a_3 ~=~ 1$$ and $ u_3~=~0$, these set of equations
reduce to a famous Euler equation or Nahm's equation
$$ \frac{dT_i}{dt} ~=~ \epsilon_{ijk} [T_j,T_k] \hbox{  } \mbox{  } i,j,k = 1,2,3, $$
where $T_is$ are $SU(2)$ generators.

{\bf Acknowledgement}: Author would like to record here the debt which the work 
owes to the papers of K. Takasaki and T. Takebe and L. Takhtajan and to many 
illuminating discussions with Professors K. Takasaki. He is also grateful
to Professors Dmitry Alekseevsky, Simon Gindikin, Warner Nahm and
Ian Strachan for valuable discussions.

\section{Volume preserving integrable systems}

 In this section we shall follow the approach of Takasaki-Takebe's [6,7]
method of studing area preserving diffeomorphic ( or sDiff$(2)$ ) K.P. equation. 
In fact, they adopted their method from self dual vacuum Einstein equation
theory. In the case of self dual vacuum Einstein equation, and hyperK\"ahler geometry 
also area preserving diffeomorphism appear, where the spectral variable is 
merely parameter. But in case of sDiff$(2)$ K.P. equation the situation is 
different, where Takasaki-Takebe showed that one has to treat $\lambda$ 
as a true variable and it enters into the definition of the Poisson bracket.  
The situation is similar here, onle the K\"ahler like two form and the associated 
"Darboux coordinates" is replaced by volume form and the Poisson bracket is 
replaced by its higher order Poisson bracket called Nambu bracket.

\bigskip
 
Suppose we consider $L = L(\lambda, p, q)$, $M = M(\lambda, p, q)$ and 
$N = N(\lambda, p, q)$ are some Laurent series in $\lambda$ with coefficients are
functions of $p$ and $q$.

\begin{defi}
The volume preserving integrable hierarchy is defined by
\bqn \frac{\p L}{\p t_n} ~=~ \{B_{1n}, B_{2n}, L\} \nqn
\bqn \frac{\p M}{\p t_n} ~=~ \{B_{1n}, B_{2n}, M\} \nqn
\bqn \frac{\p N}{\p t_n} ~=~ \{B_{1n}, B_{2n}, N \} \nqn
and \bqn \{L, M, N \} = 1, \nqn 
where $ B_{1n} :~=~ (L^n)_{n \geq 0}$ and $ B_{2n} ;~=~ (M^n)_{n \geq 0}$.
The first three equations are hierarchy equations and these constitute 
three flows of the system and the last one shows the volume preservation condition.
\end{defi}

\bigskip

Let us now compare our case with the area preserving KP hierarchy.
The sdiff(2) KP hierarchy which is given by
$$ \frac{\p {\cal L}}{\p t_n} = \{ {\cal B}_n, {\cal L} \} \hbox{    }
\mbox{    } \frac{\p {\cal M}}{\p t_n} = \{ {\cal B}_n, {\cal M} \}
\hbox{    } \mbox{   } \{ {\cal L}, {\cal M} \} = 1, $$
where ${\cal L}$ is a Laurent series in an indeterminant $\lambda$ of the form
$$ {\cal L} = \lambda + \sum_{n=1}^{\infty} u_{n+1}(t) \lambda^{-n}, $$ 
$ {\cal B}_n = ({\cal L})_{\geq 0}$. The function ${\cal M}$ is called
Orlov function and it is defined by
$$ {\cal M} = \sum_{n=1}^{\infty} nt_n {\cal L}^{n-1} + x + \sum_{i=1}^{\infty}
v_i {\cal L}^{-i-1} $$
where $t_1 = x$.

\begin{rem}
Our hierarchy has a structure of volume preserving KP hierarchy
and instead of one Orlov function we need two Orlove functions Viz.
${\cal M}$ and ${\cal N}$.
\end{rem}

Since the flows are commuting, we obtain

\begin{prop}
The Lax equation for $L$, $M$, $N$ are equivalent to the following equations:
 $$ \{ \frac{\p B_{1n}}{\p t_m},B_{2n} \} + \{ B_{1n}, \frac{\p B_{2n}}{\p t_m} \} 
 + \{ {\hat B_1}, B_{2m} \}$$ 
$$ - \{ \frac{\p B_{1m}}{\p t_n}, B_{2m} \}
- \{ B_{1m}, \frac{\p B_{2m}}{\p t_n} \} + \{ B_{1m}, {\hat H}_2 \}  = 0, $$
where $$ {\hat H}_1 = \{ B_{1n}, B_{2n}, B_{1m} \} $$ and
      $$ {\hat H}_2 =  \{B_{1n}, B_{2n}, B_{2m} \}. $$
\end{prop}

Proof: 
 Result follows from the compatibility conditions of hierarchy equations 
 and the fundamental identity.

$\Box$

\bigskip

\begin{rem}
In the case ordinary Poisson geometry and sDiff$(2)$ hierarchy this is a 
zero curvature equation.
\end{rem}

Let $\Omega$ be a three form given by
$$ \Omega :=~ \sum_{n=1}^{\infty} dB_{1n} \wedge dB_{2n} \wedge dt_{n} = 
d\lambda \wedge dp \wedge dq + \sum_{n=2}^{\infty} dB_{1n} \wedge dB_{2n} 
\wedge dt_{n}. $$
From the definition it is clear $\Omega$ is closed $3$ form. 
In fact sDiff$(3)$ structure is clearly exhibited from this structure 
and the theory is integrable in the sense of nonlinear graviton construction. 
This is a generalization of nonlinear graviton construction.

\begin{theo}
The volume preserving hierarchy is equivalent to the exterior differential equation
 $$ \Omega ~=~ dL \wedge dM \wedge dN. $$
\end{theo}

{\bf Proof}: We have seen that $\Omega$ can be written in two ways. 
Expanding both sides of the exterior differential equation as linear 
combinations of $ d\lambda \wedge dp \wedge dq $, $d\lambda \wedge dp \wedge dt_n$, 
$d\lambda \wedge dq \wedge dt_n$ and $dp \wedge dq \wedge dt_n$. \\
When we pick up the coefficients of $ d\lambda \wedge dp \wedge dq$, we obtain
the volume preserving condition
$$ \{ L, M, N \} = 1. $$

When we equate the other coefficients, viz. $ d\lambda \wedge dp \wedge dt_n$,
$d\lambda \wedge dq \wedge dt_n$ and $ dp \wedge dq \wedge dt_n$ we obtain 
the following identities:
\bqn \frac{\p ( B_{1n}, B_{2n} )}{\p (\lambda , p)} = 
\frac{\p (L, M, N)}{\p (\lambda, p, t_n)}, \nqn
\bqn \frac{\p (B_{1n},B_{2n})}{\p (\lambda, q)} = 
\frac{\p (L, M, N)}{\p (\lambda, q, t_n )} \nqn and
\bqn \frac{\p (B_{1n}, B_{2n})}{\p ( p, q)} = 
\frac{\p (L, M, N)}{\p ( p, q, t_n )} \nqn 
respectively. We multiply the above three equations by $\frac{\p L}{\p q}$,
$ \frac{\p L}{\p p}$ and $ \frac{\p L}{\p \lambda}$ respectively. 
If we add first and third equations and substract the second one from them, 
then after using the volume preserving identity we obtain,
 $$ \frac{\p L}{\p t_n} =  \{ B_{1n}, B_{2n}, L \}. $$
Similarly we can obtain the other equations also, in that case we will 
multiply the equations (20),(21) and (22) by $\frac{\p M}{\p q}$, 
$\frac{\p M}{\p p}$ and $\frac{\p M}{\p \lambda}$ respectively.

$\Box$

\bigskip

We can write down the fundamental relation
     $$ \Omega ~=~ dL \wedge dM \wedge dN $$
by \bqn  
d( M dL \wedge dN + \sum_{n=1}^{\infty} B_{1n} dB_{2n} \wedge dt_n) ~=~ 0. \nqn 

This implies the existence of one form $Q$ such that
 \bqn dQ ~=~ M d(LdN) + \sum_{n=1}^{\infty} B_{1n} d(B_{2n} dt_n). \nqn

This is an analogue of "Krichever potential" in the volume preserving case.
Hence we can say from (24)
\bqn  M ~=~ \frac{\p Q}{\p (LdN)}|_{B_{2n},t_n \hbox{ fixed }} \nqn
\bqn  B_{1n} ~=~ \frac{\p Q}{\p (B_{2n}dt_n)}|_{L,N,B_{2m},t_m ( m \neq n) 
\hbox{  fixed }}.  \nqn

\section{ Application to mulitidimesional integrable systems and 
Riemann-Hilbert problem}

We already stated that our situation is quite similar to nonlinear graviton 
construction of Penrose [11] for the self dual Einstein equation. This is a
generalization of nonlinear graviton constructions.

To the geometer self dual gravity is nothing but Ricci flat K\"ahler geometry
and it is characterized by the underlying symmetry groups $sDiff(2)$, this are 
called area preserving diffeomorphism group on surfaces. These are the natural 
generalization of the groups $Diff(S^1)$, diffeomorphism of circle.

It is well known how the area preserving diffeomorphism group appears in the self 
dual gravity equation. Let us give a very rapid description of this.

Let us start from a complexified metric of the following form
$$ ds^2 ~=~ \hbox{  det  }\left(\begin{array}{cc}
                             \matrix e^{11} &  e^{12} \\
                                     e^{21}  &  e^{22} \\
                                  \end{array}\right) $$
             $$ ~=~ e^{11}e^{22} - e^{12}e^{21}, $$
where $e^{ij}$ are independent one forms. Ricci flatness condition boils down to 
the closedness of
\bqn d\Omega^{kl} ~=~ 0 \nqn
of the exterior $2$-forms
\bqn \Omega^{kl} ~=~ \frac{1}{2} J_{ij} e^{ik} \wedge e^{jl}, \nqn
where $(J)$ is the normalized symplectic form, i.e. it is a $2 \times 2$ matrix 
whose entries are $0$, $1$, $-1$ and $0$ respectively. Then above system of two 
forms can be recast to 
\bqn \Omega (\lambda ) ~=~ \frac{1}{2} J_{ij} ( e^{i1} + e^{i2} \lambda ) 
\wedge ( e^{j1} + e^{j2}\lambda ). \nqn
This satisfies
\bqn \Omega (\lambda ) \wedge \Omega (\lambda) ~=~ 0 \nqn
\bqn d \Omega (\lambda) ~=~ 0, \nqn
where $d$ stands for total differentiation. These suggest us to introduce a pair of 
Darboux coordinates
 $$ \Omega (\lambda) ~=~ dP \wedge dQ $$
and these are the sections of the twistor fibration
   $$ \pi ~:~ {\cal T} \longrightarrow CP^1, $$
where ${\cal T}$ is the curved twistor space introduced by Penrose. Basically 
each fibre is endowed with a symplectic form and as the base point moves this
also deforms and  here comes the area preservation.

Two pairs of Darboux coordiantes are related by
$$  f (\lambda, P(\lambda), Q(\lambda) ) ~=~ P^{\prime} $$
$$  g (\lambda, P(\lambda), Q(\lambda) ) ~=~ Q^{\prime} $$
and $f$ and $g$ satisfy $ \{ f , g \} ~=~1.$ The pair $(f,g)$ is called
twistor data. Locally $f$ and $g$  ( after twisting with $\lambda$ ) yield
patching function. Ricci flat K\"ahler metric is locally encoded in this data.
This set up is nothing but the Riemann-Hilbert problem in area preserving 
diffeomorphism case.

The novelty of this approach is that this twistor construction will work in the 
higher dimensions too, when there is no twistor projection. The most important 
example is the electro-vacuum equation, volume preserving diffeomorphism groups 
in three dimension play a vital role here. This model was first introduced
by Flaherty [12] and later Takasaki [2] showed how this works explicitly.

\subsection{ Gindikin's bundle of forms}

We already stated that anti-self dual vacuum equations govern the behaviour of
complex 4-metrics  of signature $(+,+,+,+)$ whose Ricci curvature is zero and 
whose Weyl curvature is self dual. These two curvatures are independent of change 
of coordinates, so in one particular of the equations these metric becomes 
autometically K\"ahler and can be expressed in terms of a single scalar function
$\Omega$, the K\"ahler potential. Then curvarure conditions will lead you to
Ist Plebenski's Heavenly equation

\bqn \frac{\p^2 \Omega}{{\p x}{\p \tilde x}} \frac{\p^2 \Omega}
{{\p y}{\p \tilde y}} - \frac{\p^2 \Omega}{{\p x}{\p \tilde y}} 
\frac{\p^2 \Omega}{{\p y}{\p \tilde x}} = 1 \nqn

and the correspoding anti-self-dual Ricci flat metric is 
$$
g(\Omega) ~=~ \frac{\p^2 \Omega}{{\p x^i}{\p \tilde x^j}} dx^i d{\tilde x}^j,
\hbox{   } \mbox{  } {\tilde x}^i = {\tilde x}, {\tilde y}, \mbox{  } \hbox{  } x^j = x, y.
$$

This system is completely integrable and it is an example of a multidimensional 
integrable system.

\bigskip

Let $\Omega$ be the 2-form, given by
$$
\Omega = dx \wedge dy + \lambda (\Omega_{x{\tilde x}}dx \wedge d{\tilde x} +  
  \Omega_{x{\tilde y}} dx \wedge d{\tilde y} + \Omega_{y{\tilde x}} dy 
  \wedge d{\tilde x} + \Omega_{y{\tilde y}}dy \wedge d{\tilde y}) + 
  \lambda^2 d{\tilde x}\wedge d{\tilde y}.
$$

\bigskip

\begin{lem}
$$(1) \hbox{  } \mbox{  }   d\Omega ~=~ 0 $$
$$(2) \hbox{  } \mbox{   } \Omega \wedge \Omega ~=~ 0. $$
\end{lem}
 
Proof:
 Since $\Omega$ satisfies Plebenski's Heavenly Ist equation, hence $(2)$
is true.

$\Box$

\bigskip

A number of multidimensional integrable systems can be written in terms of a 
2-form $\Omega$ which satisfies the equations (4.1).

{\bf Example}:

The dispersionless KP hierarchy has a Lax representation with respect to a 
series of independent (``time'') variables $ t~=~ (t_1, t_2, \cdots )$
$$
\frac{\p {\cal L}}{\p t_n} ~=~ \{ B_n , {\cal L} \} 
$$
where
$$
 B_n := ({\cal L}^n)_{\geq 0}  \hbox{  } \mbox{   } n~=~ 1,2, \cdots
$$
${\cal L}$ is a Laurent series in an indeterminant $\lambda$ of the form

$$
{\cal L} ~=~ \lambda + \sum_{n=1}^{\infty} u_{n+1}(t)\lambda^{-n},
$$
$\{ ~,~ \}$ is a Poisson bracket in 2D phase space with respect to $(\lambda , x)$.

Let us consider
$$
\Omega ~=~ d\lambda \wedge dx + \sum_{n=2}^{\infty} dB_n \wedge dt_n
$$

then $ \Omega \wedge \Omega$ is equivalent to the zero curvature condition
\bqn
\frac{\p B_n}{\p t_m} - \frac{\p B_m}{\p t_n} +\{B_n , B_m \} ~=~ 0.
\nqn
This is an alternative form of dispersionless KP hierarchy.

All these systems are related to area preserving diffeomorphism group sDiff(2).

\bigskip

In this paper we are presenting an analogous picture for multidimensional  
integrable systems related to volume preserving diffeomorphism group.

\begin{prop}
 The 3-form 
$$
\Omega^{(3)} ~=~ d\lambda \wedge dp \wedge dq + \sum_{n=2}^{\infty}dB_{1n} \wedge
dB_{2n} \wedge dt_n,
$$
 satisfies

$$ d\Omega^{(3)} = 0 $$
$$ \Omega^{(3)} \wedge \Omega^{(3)} ~=~ 0. $$
\end{prop}

This is equivalent to propostion 3.2. Hence it is satisfied.

$\Box$

The Gindikin's method [13] of pencil of 2-forms is the  most effective way to study 
these systems. Consider the following system of Ist order equation depending on a 
parameter 
$ \tau ~=~ (\tau_1 ,\tau_2 ) \hbox{  } \in {\bf C}^2 $

$$ e^1 (\tau) ~=~ e^{11} \tau_{1}^{k} + \cdots + e^{1k}\tau_{2}^{k}, $$
$$ e^2 (\tau) ~=~ e^{21} \tau_{1}^{k} + \cdots + e^{2k}\tau_{2}^{k}, $$
$$...................................................$$
$$.................................................. $$
$$ e^{2l}(\tau) ~=~ e^{2l1}\tau_{1}^{k}+ \cdots + e^{2lk}\tau_{2}^{k}. $$

where $e^{ij}$ are $1$-forms. Let $\Omega^k (\tau)$ be the bundle of 2-forms

$$ \Omega^k (\tau) ~=~ e^1 (\tau) \wedge e^2 (\tau) + \cdots + e^{2l-1}(\tau) 
\wedge e^{2l}(\tau) $$
satisfying the conditions
$$ \bullet (\Omega^k )^{l+1} ~=~ 0 $$
$$ \bullet (\Omega^k )^l ~\neq~ 0 $$
$$ \bullet d\Omega^k ~=~ 0. $$
 The bundle of forms actually encodes the integrability of the original system. 
 In the special case $l~=~1, \hbox{  } k~=~2$, we recover the Ricci flat 
 metric

$$ g ~=~ e^{11}e^{22} - e^{12}e^{21}. $$

\subsection{ Higher dimensional analogue of Gindikin's Pencil}

Let us consider the following system of Ist order equations depending
on parameter $$ \tau ~=~ (\tau_1, \tau_2, \tau_3)  \in  {\bf C}^2 , $$

$$ e^1 (\tau ) ~=~ e^{11}\tau_1 + e^{12}\tau_2 + e^{13}\tau_3, $$
$$ e^2 (\tau ) ~=~ e^{21}\tau_1 + e^{22}\tau_2 + e^{23}\tau_3, $$
$$ e^3 (\tau ) ~=~ e^{31}\tau_1 + e^{32}\tau_2 + e^{33}\tau_3. $$

Just like the previous situation these bundle of forms then encodes the        
integrability of the original system. The metric defined here is the higher 
order analogue of Ricci metric in two form case.

This metric is given by,

$$ g ~=~ e^{11}e^{22}e^{33} - e^{11}e^{32}e^{23} + e^{12}e^{31}e^{23} $$
$$ - e^{12}e^{21}e^{33} + e^{13}e^{21}e^{32} - e^{13}e^{31}e^{22}. $$

\begin{rem}

Since 3-form $\Omega^{(3)}$ satisfies 
$$ d\Omega^{(3)} ~=~ 0 $$
$$ \Omega^{(3)} \wedge \Omega^{(3)} ~=~ 0, $$
so it denotes the volume preserving multidimensional integrable systems. These
  systems can be described by Gindikin's bundle of multi forms, higher 
dimensional analogue of nonlinear graviton. 
\end{rem}

\subsection{Twistor description of Volume preserving multidimensional integrable 
systems}

The natural question would be to find out the analogous Riemann-Hilbert problem 
in the volume preserving case. Our situation is very similar to electro-vacuum equation. 
Let us consider two sets of solutions of hierarchy $( L, M, N)$ and 
$({\hat L}, {\hat M}, {\hat N})$ with different analysity. 
Then there exist an invertible functional relation between these 
two sets of functions such that it satisfies
 $$ {\hat L} ~=~ f_1 ( L, M, N),   \hbox{      } \mbox{     } {\hat M} 
 ~=~ f_2 ( L, M, N ),$$
  $$  {\hat N} ~=~ f_3 (L, M, N), $$
where $ f_1 = f_1 (\lambda, p, q)$, $ f_2 = f_2 (\lambda, p, q)$ and 
$ f_3 = f_3 (\lambda, p, q )$ are arbitrary holomorphic functions defined 
in a neighbourhood
of $\lambda = \infty$ except at $ \lambda = \infty$. 

We assume $f_1, f_2, f_3 $ satisfy the canonical Nambu Poisson relation
   \bqn \{ f_1, f_2, f_3 \} ~=~ 1. \nqn

This is a kind of Riemann-Hilbert problem related to three dimensional diffeomorphisms. In this case sDiff$(3)$ symmetries is clear, in fact sDiff$(3)$ group acts on $( f_1, f_2, f_3 )$, we can lift this action on $ (L, M, N)$ and $({\hat L}, {\hat M}, {\hat N})$ via Riemann-Hilbert fatorization. 

\subsection{ Application to hydrodynamic type systems}

There are certain kind of integrable systems, called hydrodynamic type,
naturally arise in gas dynamics, hydrodynamics, chemical kinematics and may other 
situations, these are given by

\bqn u_{t}^{i} ~=~ v_{j}^{i}(u) u_{x}^{j}, \nqn
where $v_{j}^{i}(u)$ is an arbitrary $ N \times N$ matrix function of 
$$ u ~=~ (u^1, \cdots ,u^N), \hbox{   } \mbox{  }  u^i ~=~ u^i(x,t) \hbox{  }
i~=~1, \cdots ,N. $$

The Hamiltonian systems of hydrodynamic type systems considered above
have the form
\bqn u_{t}^{i} ~=~ \{u^i~,~H\},\nqn
where $H ~=~ \int h(u)dx$, is a functional of hydrodynamic type. 
The Poisson bracket of these systems has the form
\bqn \{ u^i(x)~,~u^j(y) \} ~=~ g^{ij}(u) \delta_x (x-y) + 
b_{k}^{ij}(u) u_{x}^k \delta (x-y), \nqn
called Dubrovin-Novikov type Poisson bracket. Dubrovin-Novikov showed if the
metric is non degenerate i.e. det $[g^{ij}] ~\neq 0$
then (37) yields a Poisson bracket provided $g^{ij}(u)$ is a metric of zero 
Riemannian Curvature.

 A large class of them can be described by the Lax form,
$$ \psi_x ~=~ zA\psi $$
$$ \psi_t ~=~ zB\psi. $$

From the compatibility condition we obtain,
$$ A_t ~=~ B_x $$
$$ AB ~=~ BA. $$
These set of equations can be easily recasted to

$$ d\Omega ~=~ 0 $$
$$ \Omega \wedge \Omega ~=~ 0, $$
where $\Omega$ is a closed one form.

Hence has also a twistorial description. The upshot of this section is that
Gindikin pencil of forms can be applied to large number of classes integrable
systems, including the volume preserving integrable systems we propose here.

\bigskip

\section{Conclusion}

In this paper we have shown that there are large class of integrable systems
can be obtained from Nambu-Poisson mechanics. They belong to the same family
of self dual Einstein or dispersionless KP type equations. In fact they
are related to what should be called volume preserving KP system. In some sense these 
integrable systems are higher dimensional generalization of self dual
Einstein equation. Hence these systems are describable via Gindikin's
bundle of forms, or twistor method.

In fact we have pointed out in this paper that Nambu-Poisson manifold is an 
useful tool to study volume preserving integrable systems. 
Recently in membrane theory physicists [14] have
found M-algebra from M-brane, these are related to Nambu-Poisson mechanics.

\bigskip

There are certain problems we have not discussed in this paper, viz. the 
quantization of these volume preserving generalized multidimensional
integrable systems. Presumably method of star product quantization [15,16] would
be the best way quantize these systems and
instead of binary star product we need triple star
product [17,18]. Thus it is important to see whether Moyal-Nambu bracket
can be expressed in terms of Fedosov's triple star product.

\section{References}

1. Guha, P. {\em Volume preserving integrable systems and Nambu-Poisson
manifolds.} Class. Quant. Gravity {\bf 14}, L37-L43 (1997).\\
2. Takasaki, K. {\em Volume preserving diffeomorphisms in integrable 
deformation of self-dual gravity} Phys. Lett.{\bf B285}, 187-190 (1992). \\
3. Nambu, Y., {\em Generalized Hamiltonian Dynamics} 
Phys.~Rev.~D. {\bf 7}, 2405 (1973).\\
4. Takhtajan, L.A., {\em On Foundation of the Generalized Nambu Mechanics}
 Comm.~Math.~Phys. {\bf 160}, 295 (1994). \\
5. Chatterjee, R, Takhtajan, L.A. {\em Aspects of Classical and Quantum Nambu
Mechanics} Lett.~Math.~Phys. {\bf 37}, 475 (1996).\\
6. Takasaki, K. and Takebe, T. {\em SDiff(2) KP hierarchy.}
Int. J. Mod. Phys. {\bf 7A}, 889 (1992). \\
7. Takasaki, K. and Takebe, T. {\em Integrable hierarchies and dispersionless
 limits. } preprint, UTMS 94-35.\\
8. Alekseevsky, D.V. and Guha, P. {\em On decomposibility of Nambu tensor.} 
Acta Math. Com. Univ. {\bf 65}, 1 (1996). \\
9. Marmo, G, Vilasi, G. and  Vinogradov A.M. {\em The local
structure of n-Poisson and n-Jacobi manifolds} J. Geom. Phys. {\bf 25}, 141 (1998).\\
10. Bloch, A.M. and Marsden, J.E. {\em Stabilization of rigid body 
dynamics by the energy-Casimir method.} 
Systems and control letters. {\bf 14}, 341 (1990).\\
11. Penrose, R.P. {\em Nonlinear gravitons and curved twistor theory.} Gen.
Rel. Grav. 7, {\bf 31} (1976).\\
12. Flaherty, E.J. {\em The nonlinear graviton in interaction with photon.}
Gen. Rel. Grav. {\bf 9}, 961 (1978). \\
13. Gindikin, S.G. {\em Generalized conformal structures} 
Twistors in Mathematics and Physics. ed. T.N. Bailey and 
R.J. Baston, LMS lecture note series {\bf 156}, CUP, Cambridge. \\
14. Hoppe, J. {\em On M-algebras, the quantization of Nambu-mechanics and
 volume preserving diffeomorphisms.} hep-th 9602020, ETH-TH/95-33.\\
15. Bayen, F., Flato, M., Fronsdal, C., Lichnerowicz, A., Sternheimer, D.,
 {\em Deformation Theory and Quantization:I. Deformations of Symplectic
 Structures.} Ann.~Phys. {\bf 110}, 67 (1978). \\
16. Kontsevich, M. {\em Deformation Quantization of Poisson Manifolds.}
Preprint IHES/M/97/72; q-alg/9709040 \\  
17. Strachan, I.A.B. {\em A Geometry for Multidimensional Integrable Systems}
Jour. Geom. Phys. {\bf 21}, 255 (1997).\\
18. Dito, G, Flato, M, Sternheimer, D. and Takhtajan, L. 
{\em Deformation Quantization and Nambu Mechanics} Comm. Math. Phys. {\bf 183}, 
1 (1997).\\

\end{document}